\documentclass[twocolumn,aps,prb,superscriptaddress,longbibliography,floatfix, reprint]{revtex4-1}

\usepackage{natbib}
\usepackage{times}
\usepackage{graphicx}
\usepackage{amsfonts}
\usepackage{amsmath,amsthm,amssymb,mathrsfs}
\usepackage{dsfont}
\usepackage{xcolor}
\usepackage{microtype}
\usepackage{bm}
\usepackage{siunitx}
\usepackage[colorlinks=true, allcolors=black, citecolor=blue, linkcolor=blue, urlcolor=blue]{hyperref}

\usepackage{float} 
\usepackage{subfigure}
\usepackage{booktabs}

\allowdisplaybreaks

\DeclareMathOperator{\diag}{diag}

\newcommand{\mean}[1]{\big<{#1}\big>}

\newcommand{\su}{\uparrow}
\newcommand{\sd}{\downarrow}

\usepackage[capitalize]{cleveref}

\begin{document}

\title{Bending-strain effects in conventional superconductors and superconducting junctions}

\author{Kjell S. Heinrich}
\author{Henning G. Hugdal}
\author{Morten Amundsen}
\author{Sol H. Jacobsen$^{*,}$}
%\thanks{Corresponding author: sol.jacobsen@ntnu.no}
\affiliation{Center for Quantum Spintronics, Department of Physics, NTNU, Norwegian University of Science and Technology, NO-7491 Trondheim, Norway\\$^*$Corresponding author: sol.jacobsen@ntnu.no}

% \date{\today}

\begin{abstract}
    We consider the effect of bending-strain in thin films of clean, conventional superconductors (S), and the proximity-induced effect of this strain in SN bilayers with a normal metal (N), and SNS junctions with equal curvatures in each superconductor. We find that the effective spin-orbit coupling due to strain in the superconductor induces both spin-polarized and unequal-spin even-frequency p-wave triplet pairings throughout the superconductor. When interfaced with a normal metal, additional odd-frequency pairings are induced, and their magnitudes can be tuned with the strain. In SNS junctions, the strain alone can induce a superconducting spin current in the junction. The spin-polarized current can undergo a $0-\pi$-transition, resulting in an in-plane, strain-induced magnetization that switches sign as a function of the strain. We discuss the underlying physics and its implications for superconducting spintronics.
\end{abstract}

\maketitle

\section{Introduction}\label{Sec:Intro}
Conventional superconductors exist in plentiful elemental or alloyed form, and have been studied for decades, but their response to different stimuli -- on their own or in conjunction with other materials -- continue to provide novel fundamental insights. Deforming a superconducting wire or film results in bending-strain, and sometimes torsion, and will typically degrade the critical current that the superconductor can support, up to some critical value of strain beyond which the superconductivity breaks down altogether \cite{Kaiho1980,Clem2011,Allen_2014}. However, recent studies have shown that bending-strain, sometimes referred to as curvature-induced strain in the literature, can also have profound effects on the superconductivity when applied in a material adjacent to the superconductor, through its modulation of the proximity effect \cite{Salamone2021curvature,salamone2022curvature,skarpeid2024non,Salamone2024Interface,Salamone2024spin_valves}. These studies have shown that the effective spin-orbit coupling (SOC) introduced by curvature is instrumental in converting conventional s-wave, even-frequency, spin-singlet correlations into spin-polarized triplet correlations. Achieving spin-polarized superconducting correlations from abundant and versatile materials is highly desirable in order to, for example, combine zero-resistance currents with spin-manipulation for spintronic devices \cite{Linder2015, Eschrig2011,Eschrig2015,Amundsen2024}.

Modulation of the singlet order parameter occurs under the effect of SOC, an external magnetic field or an internal exchange field. 
When a superconductor is proximity-coupled to a ferromagnet or antiferromagnet, or exposed to an external magnetic field, odd-frequency, s-wave \textit{spin-polarized triplet} correlations can be generated. That is, the generation of these correlations requires both spin-splitting and a second spin-orienting field, typically engineered via magnetic multilayers or conical exchange magnets \cite{Eschrig2011,Robinson2010,Khaire2010}, or by employing intrinsic SOC (for example from a heavy metal) in combination with a single exchange field \cite{Bergeret2013,Bergeret2014,Jacobsen2015b}. Recent studies have shown that geometric curvature, increasingly harnessed for its advantages in spintronics \cite{Gentile2022nanoscale_curved,Ortix2015Quantum_torsion,gentile2013strain,Ying2016spin_texture,Das2019Curved_spintronics,makarov2022new,fomin2022perspective,Kopasov2021}, is an effective way to both design and manipulate curvature-induced SOC. Geometric curvature alone is therefore sufficient to affect the symmetries of superconducting correlations persisting in single magnetic elements \textit{i.e. without using any intrinsic SOC, magnetic multilayers or strain} \cite{Salamone2021curvature,salamone2022curvature,skarpeid2024non,Salamone2024Interface,Salamone2024spin_valves}. The addition of bending-strain may then be used to manipulate the different superconductivity populations in-situ, for example to control the superconducting transition \cite{salamone2022curvature,Salamone2024spin_valves}, or the direction of Josephson current \cite{Salamone2021curvature,skarpeid2024non}. 

In contrast to its role in magnetic materials, geometric curvature \textit{alone} (i.e. curvature without strain) is not expected to affect the symmetries of the order parameter of a plain s-wave superconductor (S). With ballistic transport, an interface with a normal metal (N) will convert a portion of the even-frequency, s-wave singlet correlations into odd-frequency p-wave singlets \cite{Tanaka2007OddInterface}. When proximity-coupled to a non-magnetic material such as a semiconductor, normal metal or topological insulator, geometric curvature in the non-magnetic material can then affect the singlet order parameter in the presence of SOC, due to its tunable effect on the generation of unpolarised triplet correlations \cite{Ying2017Tuning, Kutlin2020}. For example, curvature in a proximity-coupled topological insulator, which has strong SOC, can induce topological superconductivity \cite{Chou2021}. Spin-orbit-coupled superconducting wires have also been found to display the evolution of topological phases themselves due to geometric deformation \cite{Francica2020Topological_phases}. Bending-strain introduces an effective SOC with a rotating normal vector. While of Rashba-type, the rotating angle affects the generation and modulation of triplet correlations differently from a constant, intrinsic Rashba-SOC, or extrinsic SOC from gating. In this article, we explore the extent to which bending-strain affects the symmetries of the order parameter in a conventional, even-frequency s-wave spin-singlet superconductor alone, and we examine the implications of the symmetry transformations on the proximity effect with normal metals in bilayers and Josephson junctions.

\begin{figure}[hbt]
    \centering
    \includegraphics[width = 0.48\textwidth]{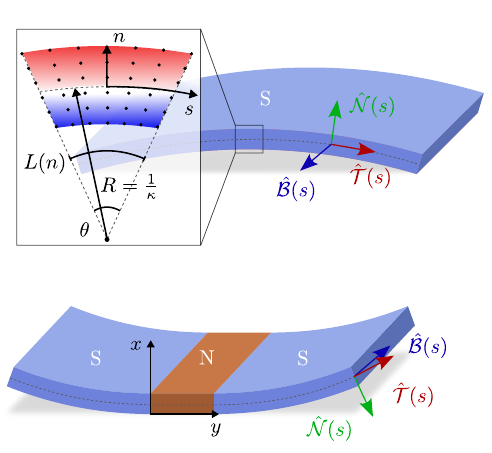}
    \caption{\label{fig:curvilinear coordinates} Thin-film superconductor with bending-strain along a constant curve with arclength $s$ (top), and (below) an SNS Josephson junction with uniformly curved superconductors and an uncurved normal metal. Orthonormal, curvilinear unit vectors in the tangential $\hat{\mathcal{T}}(s)$, normal $\hat{\mathcal{N}}(s)$, and binormal $ \hat{\mathcal{B}}(s)$ direction, as well as the radius of curvature $R$, are indicated. In the inset, red represents tensile strain and blue indicates compressive strain of the lattice.}
\end{figure}

We will use the Bogoliubov-de Gennes framework \cite{zhu2016bogoliubov} to show how the effective SOC of bending-strain in clean, conventional superconductors (see \Cref{fig:curvilinear coordinates}) can modulate the symmetries of the order parameter, generating even-frequency p-wave triplets, both with and without spin polarization. We will show how this combines with symmetry-modulations from normal-metal interfaces in SN bilayers and SNS Josephson junctions with a straight normal metal: In bilayers, the magnitude of the additional odd-frequency pairings can be tuned with the strain, and we find an in-plane, strain-induced magnetization that switches sign as a function of the strain in SNS junctions.

\section{Results}\label{Sec:Results}
The order parameters and observables we will examine are defined and derived in curvilinear coordinates in \cref{SubSec:OrderObs}. In \cref{Subsec:Results S}, we explore the effects of bending-strain in clean, conventional superconducting films. In \cref{Subsec: SN}, we consider the additional effect of proximity-coupling the strained superconductor to a normal metal, and in \cref{Subsec:Results SNS} we consider the proximity-induced effects in SNS junctions. The theoretical details, including how strain-induced SOC enters in the curvilinear framework, as well as the discretization and diagonalization of the continuum Hamiltonian, are provided in the Methods, \cref{Sec:Methods}. Parameter regimes for the numerical studies are explained and illustrated in \cref{SubSec:Methods Numerical}.

\subsection{Order parameters and observables}\label{SubSec:OrderObs}
To compare different superconducting symmetries, we define the order parameters as $\Delta_{\bm{i}} (\tau) = \langle \mathcal{T} c_{\bm{i}, \sigma}(\tau)\: c_{\bm{i}', \sigma'}(0) \rangle $, where $\mathcal{T}$ is the time ordering operator, and $\tau$ is the relative time coordinate scaled by $\hbar/t$. The hopping amplitude between nearest neighbours is denoted by $t$, and $c_{\bm{i}, \sigma}(\tau)$ are the fermion annihilation operators for spin $\sigma$ at lattice site $\bm{i}$ [see Methods \cref{SubSec:Methods discretization}, for details]. Thus, we can investigate the order parameters of even- ($\tau=0$) and odd-frequency ($\tau\neq0$) correlations. In \cref{Tab:SPOT}, we introduce the notation for the relevant combinations of underlying Spin-Parity-Orbital-Time symmetries that ensure overall odd order parameter symmetry \cite{LinderBalatsky2019OddFrequency}.
\begin{table}[H]
    \centering
    \begin{tabular}{l c c c c l l}
     & \hphantom{-}$S$ & \hphantom{-}$P^*$ & \hphantom{-}$O$ & \hphantom{-}$T^*$ & $\quad$ & \\
    $\mathcal{S}_0$ & -1 & \hphantom{-}1 & \hphantom{-}1 & \hphantom{-}1 & & even-frequency $s$-wave singlet \\
    $\mathcal{S}_{\sigma\sigma'}(\tau)$ & \hphantom{-}1 & \hphantom{-}1 & \hphantom{-}1 & -1 & & odd-frequency $s$-wave triplets \\
    $\mathcal{P}^{\bm{m}}_{\sigma\sigma'}$ & \hphantom{-}1 & -1 & \hphantom{-}1 & \hphantom{-}1 & & even-frequency $p^{m}$-wave triplets \\
    $\mathcal{P}^{\bm{m}}_0(\tau)$ & -1 & -1 & \hphantom{-}1 & -1 & & odd-frequency $p^{m}$-wave singlet\\
    \end{tabular}
    \caption{Spin-parity-orbital-time symmetries of order parameters defined in the main text. The operators $T^*$ and $P^*$ are denoted with an asterisk as they only commute the relative coordinates of the electron operators and do not invert the full spaces.}\label{Tab:SPOT}
\end{table}

Adhering to the symmetry relations summarized in \cref{Tab:SPOT}, we can define the s- and p-wave singlet and triplet amplitudes as
\begin{align} 
    \mathcal{S}_{\bm{i}, 0} &= \frac{1}{2} \big[\mean{c_{\bm{i},\su} c_{\bm{i},\sd}} - \mean{c_{\bm{i},\sd} c_{\bm{i},\su}}  \big],\label{eq:definitions s0}\\
    \mathcal{S}_{\bm{i}, \su\sd}(\tau) &= \frac{1}{2} \big[\mean{c_{\bm{i},\su}(\tau)\: c_{\bm{i},\sd}(0)} + \mean{c_{\bm{i},\sd}(\tau)\: c_{\bm{i},\su}(0)} \big],\\
    \begin{split}
        \mathcal{P}_{\bm{i}, 0}^{\bm{m}}(\tau) &= \sum_{\pm}\!\pm  \frac{1}{2} \big[\mean{c_{\bm{i},\su}(\tau)\: c_{\bm{i}\pm\bm{m},\sd}(0)} \\[-5pt]
        &\qquad\qquad - \mean{c_{\bm{i},\sd}(\tau)\: c_{\bm{i}\pm\bm{m},\su}(0)} \big],
    \end{split} \\
   \mathcal{P}^{\bm{m}}_{\bm{i}, \su\sd} &= \sum_\pm\! \pm \frac{1}{2} \big[\mean{c_{\bm{i},\su} c_{\bm{i}\pm\bm{m},\sd}} + \mean{c_{\bm{i},\sd} c_{\bm{i}\pm\bm{m},\su}}  \big], \label{eq:definitions p_t0}\\
    \mathcal{P}^{\bm{m}}_{\bm{i}, \sigma\sigma} &= \sum_\pm \!\pm \mean{c_{\bm{i},\sigma} c_{\bm{i}\pm\bm{m},\sigma}}.\label{eq:definitions p_sigma}
\end{align}
Here $\bm{m}$ is one lattice spacing in either the tangential or binormal direction (see \cref{fig:curvilinear coordinates}), and the time-dependent operators are given by $ c_{\bm{i},\sigma}(\tau) = e^{i\mathcal{H}\tau} c_{\bm{i}, \sigma} e^{-i\mathcal{H}\tau}$. Observe here that the p-wave correlations have \(\bm{i} - \bm{i}' \neq 0\), and the s-wave correlations have \(\bm{i} - \bm{i}' = 0\).

The gap equation in the mean-field approximation is given by $\Delta_i = U_i\:  \mathcal{S}_{i,0}$ and can be written as
\begin{align} \label{eq:gap_bdg}
    \Delta_{i}\! =\! \frac{U_{i}}{2 N_b}\!\! \sum_{n, k}^{'}\! \Big[&x^*_{i, n, k} u_{i, n, k}\!-\!w^*_{i, n, k} v_{i, n, k} \Big]\!\tanh(\beta \mathcal{E}_{n, k}),    
\end{align}
where we have introduced a new quasiparticle basis $\gamma_{i,n,k}$ through a Bogoliubov transformation with complex coefficients $\{u_{i,n,k},v_{i,n,k},w_{i,n,k},x_{i,n,k}\}$ [details in Methods, \cref{Sec:Methods}]. Furthermore, we have used that ${\mean{\gamma^\dag_{n,k} \gamma_{n', k}} = f(2 \mathcal{E}_{n, k}) \delta_{n,n'}}$, where $\mathcal{E}_{n,k}$ is the eigenenergy associated with $\gamma_{i, n,k}$ and $f(2\mathcal{E}_{n, k})$ is the Fermi-Dirac distribution, and $\beta$ is the inverse temperature. The corresponding expressions for \cref{eq:definitions s0}-(\ref{eq:definitions p_sigma}) in the new quasiparticle basis [\cref{eq:old operators using uvwx}] that are used for computation are given in Sec. S.I., in the Supplementary Information.

We can consider the derivatives of the odd-frequency order parameters to eliminate the dependence on the relative time coordinate. The odd-frequency order parameters are odd in time and vanish at \(\tau = 0\), while their derivatives are even and typically finite when evaluated at this point \cite{AbrahamsBalatsky1995OddFreq}. Therefore, we will evaluate all correlations at $\tau = 0$. The relative time derivative is defined as  $\dot\Delta_i(\tau) = \langle \mathcal{T} \frac{\partial c_{\bm{i}, \sigma}(\tau)}{\partial \tau} c_{\bm{i}', \sigma'}(0) \rangle$, where the derivative is computed by commuting the electron operator with the Hamiltonian,
\begin{align} 
    \frac{\partial c_{\bm{i}, \sigma}(\tau)}{\partial \tau} = i \left[\mathcal{H},  c_{\bm{i}, \sigma}(\tau) \right].
\end{align}

To find the spin-current, one can consider continuity equations. Using the Heisenberg equation of motion to rewrite the time derivative, the continuity equation for the spin-current can be expressed as
\begin{equation} \label{eq:spin_curr_commutation}
    \sum_m \bm{J}_{\bm{i}, m} = - i \big[\mathcal{H},\: \bm{S}_{\bm{i}} \big].
\end{equation}
Here, $\mathcal{H}$ is the Hamiltonian and $m$ enumerates the surface normals of the unit cell $\bm{i}$. By comparing the current flowing in and out from a single unit cell in the tangential direction, a final expression can be obtained \cite{Risinggaard2019}. Thus, $\bm{J}$ denotes the spin-current flowing parallel to the tangential unit vector $\hat{\mathcal{T}}$ in real space, with components in spin space $J^\mu$, where $\mu=(s,n,b)$ indicates spin alignment in the tangential, normal, and binormal directions. The components of the spin-current $ \bm{J}$ are only conserved in areas without spin-orbit coupling. We will therefore only examine spin-current as a useful observable in cases where the material is not curved at $\bm{i}$, i.e. in an uncurved normal metal sandwiched between two curved superconductors. Applying the standard quasiparticle operators and coefficients [\cref{eq:eigenvectors} in Methods, \cref{Sec:Methods}], we find the tangential- and binormal-components of the spin-current
\begin{widetext}
\begin{align}
    \begin{split}
        J^{s}_{i}  = \frac{-2t}{N_b} \sum_{n,k, \pm}^{'} \pm \mathfrak{Re}\biggl\{(v^*_{i\pm1,n,k} u_{i,n,k} - u^*_{i\pm1,n,k} v_{i,n,k}) f(2\mathcal{E}_{n,k}) - (x^*_{i\pm1, n,k} w_{i,n,k} - w^*_{i\pm1,n,k} x_{i,n,k} ) f(-2\mathcal{E}_{n,k}) \biggr\}, \\
    \end{split}\\
    \begin{split}
        J^{b}_{i}  = \frac{-2 t }{N_b} \sum_{n, k, \pm}^{'} \pm\mathfrak{Im} \biggl\{(u^*_{i\pm1, n, k} u_{i, n, k} - v^*_{i\pm1, n, k} v_{i, n, k} ) f(2 \mathcal{E}_{n, k}) - (w^*_{i\pm1, n, k} w_{i, n, k} - x^*_{i\pm1, n, k} x_{i, n, k} ) f(-2 \mathcal{E}_{n, k}) \biggr\}.\\
    \end{split}
    \end{align}
\end{widetext}

The spin magnetization is defined as $\bm{S}_{\bm{i}} = \mean{c^\dag_{\bm{i},\alpha} \bm{\sigma}^{\alpha\beta} c_{\bm{i}, \beta}}$, which means that the binormal component becomes
\begin{align}
\begin{split}
    S^b_i = \frac{2}{N_b} \sum_{n, k}^{'}& \biggl[\bigl(|u_{i,n,k}|^2 - |v_{i,n,k}|^2\bigr) f(2\mathcal{E}_{n,k})\\
    +& \bigl(|w_{i,n,k}|^2 - |x_{i,n,k}|^2\bigr) f(-2\mathcal{E}_{n,k}) \biggr].
\end{split}
\end{align}
Having defined the order parameters and observables of interest, we can now examine their variation with increasing, uniform strain.

\subsection{Bending-strain in a thin-film superconductor}\label{Subsec:Results S}

\begin{figure}[bht]
    \includegraphics[width=0.48\textwidth]{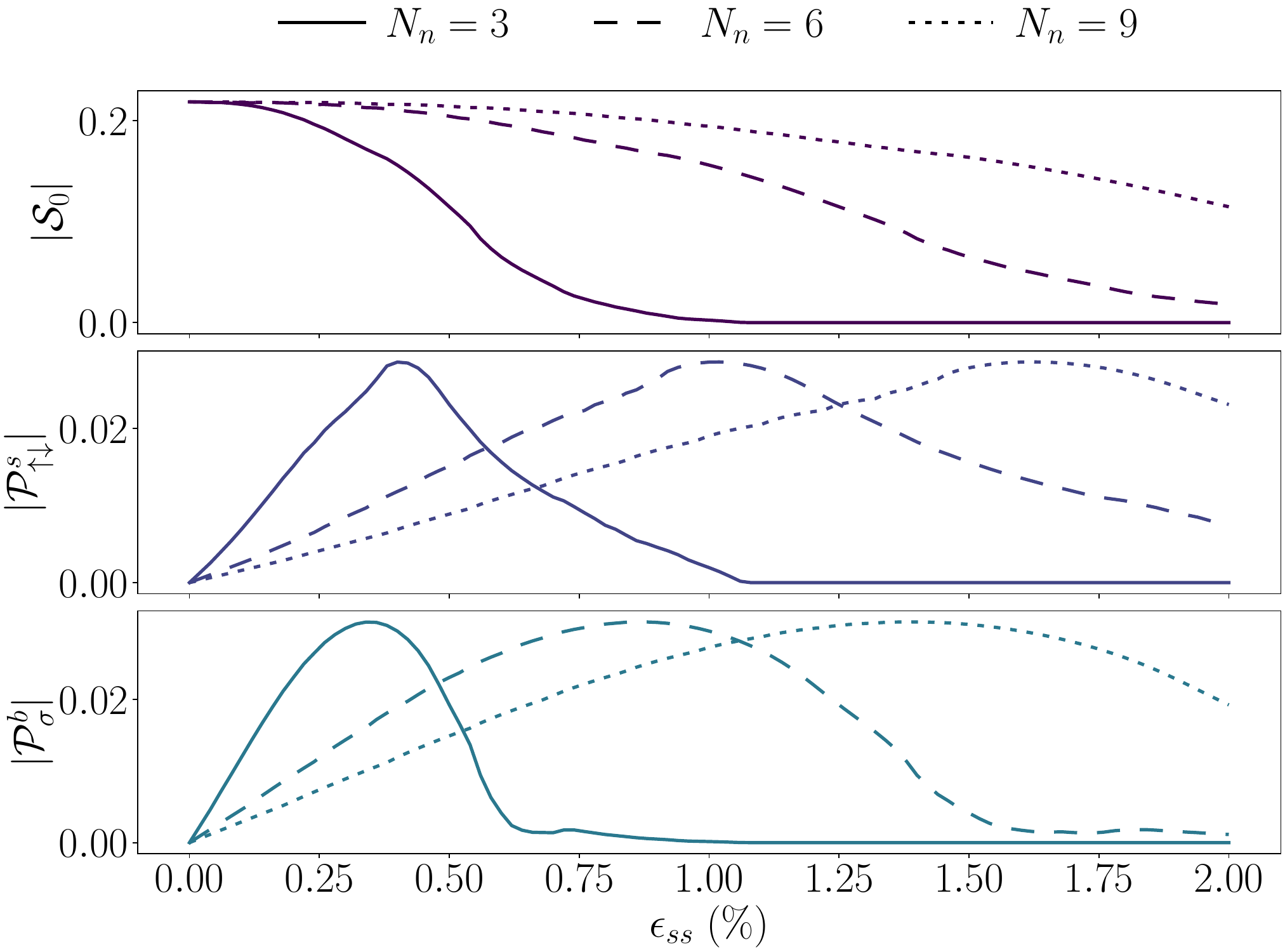}
     \caption{The even-frequency s-wave singlet and p-wave triplet pairings at the center of the superconductor as a function of the strain $\epsilon_{ss}$. The spin-polarized p-wave amplitudes are equal in magnitude $|\mathcal{P}^b_\sigma|=|\mathcal{P}^b_{-\sigma}|$. Parameters are given in the main text.}
    \label{fig:amplitudes_superconductor}
\end{figure}

The superconductor is initialized with an s-wave gap on a square lattice, and the gap equation is solved self-consistently by iterative diagonalization (details of the numerical implementation are provided in the Methods, \cref{SubSec:Methods Numerical}). Bending the superconducting film creates a strain-induced spin-orbit coupling, where the direction of asymmetry is in the normal direction to the surface. This creates an additional term in the Hamiltonian described by
\begin{align}
    {\displaystyle\mathcal{H}_{\text{soc}} = \alpha_N \big( \sigma_B \hat{k}_s - \sigma_T \hat{k}_b\big),}
\end{align}
where $\alpha_N$ is proportional to the local curvature, $\sigma_{\mu}$ are curvilinear Pauli-matrices, and $\hat{k}_\mu$ are momentum operators. This strain-induced SOC creates a mixed parity s- and p-wave order parameter in the superconductor. The magnitude of the s-wave gap will diminish as a function of the strain, as shown in Fig.~\ref{fig:amplitudes_superconductor}. This is analogous to doing self-consistency calculations for an increasing Rashba coefficient in a spin-orbit coupled superconductor, which yields the same results \cite{Ying2017Tuning}. 

We denote the p-wave triplet pairings with no spin projection as $\mathcal{P}^s_{\su\sd}$, and the spin-polarized triplets as $\mathcal{P}^b_\sigma$, as indicated in \cref{Tab:SPOT}. With time-inversion symmetry conserved, the spin-polarized p-wave amplitudes are equal in magnitude $|\mathcal{P}^b_\su|=|\mathcal{P}^b_\sd|$. The superconducting gap closes completely for high enough strain, so that both s- and p-wave correlations become negligible. The thickness of the film denoted as $N_n$, determines the maximum strain and the extent of superconductivity it can support. We assume that the strength of the SOC is proportional to the curvature, which means that a thinner film can accommodate greater curvature for the same surface strain. The curvature at a given strain is less for thicker films, and so the effective SOC is weaker. 
The singlet s-wave and finite triplet p-wave order pairings are shown in Fig.~\ref{fig:amplitudes_superconductor} as a function of the strain. While the curvature is in the tangential direction, the pairings $\mathcal{P}^s_{\uparrow\downarrow}$ and $\mathcal{P}^b_{\sigma}$ experience the same strain-induced electric field in the normal direction, leading to their similar profile in Fig.~\ref{fig:amplitudes_superconductor}. Their spin polarization depends on the chosen spin axis [c.f. \cref{eq:cont_hamiltonian}] and, in our example, both the spin axis and $\hat{\mathcal{B}}$ are aligned parallel to $\hat{z}$. The pairings arise through the entire superconductor because the superconductor has uniform circular curvature, giving a constant bending-strain profile and equivalent SOC-strength. If the strain is only applied to a certain part, the p-wave amplitudes will only be finite locally. For instance, in an ellipse, the correlations are localized at the apexes\cite{heinrich2024curvature}. The emergence of $\mathcal{P}^s$-wave correlations under bending strain aligns with existing intuition from one-dimensional systems. However, in thin films, the inclusion of the binormal direction and the term $\sigma_T \hat{k}_b$, which here leads to the emergence of $\mathcal{P}^b$-wave triplet states in strained superconducting films, is crucial to the appearance of novel effects in conjunction with interfacial symmetry breaking and Josephson currents, as we will now explore.

\subsection{SN bilayer with bending-strain in S}\label{Subsec: SN}
We consider a superconductor-normal metal (SN) bilayer, with an example of arbitrarily chosen length of superconductor with $N_S=50$ sites, a normal metal with $N_N = 21$ sites, and $N_b = 71$ sites in the binormal direction. Choosing an odd number simplifies the numerical algorithm at the edge of the Brillouin zone, but does not impact the results (see \cref{SubSec:Methods Numerical}). The film thickness in the normal direction is in this case set to $N_n = 3$, and the strain is only applied to the superconductor. 

Without curvature, the interface between the conventional, even-frequency s-wave spin-singlet (finite $\mathcal{S}_0$) superconductor and a normal metal breaks translational symmetry, which will create odd-frequency p-wave singlets [$\mathcal{P}^s_0(\tau)$]\cite{Tanaka2007OddInterface}. These p-wave singlets are dependent only on Andreev reflections \cite{Cayao2018InterfaceRashba}. The addition of bending-strain will then reduce the gap within the superconductor and create an order parameter with an even frequency mixed parity $\mathcal{S}_0$ + $\mathcal{P}^s_{\su\sd}$ and $\mathcal{P}^b_\sigma$, as seen in \cref{fig:amplitudes_superconductor}, which now penetrates into the normal metal through the proximity effect. 

\begin{figure}[htb]
    \centering    \includegraphics[width=0.48\textwidth]{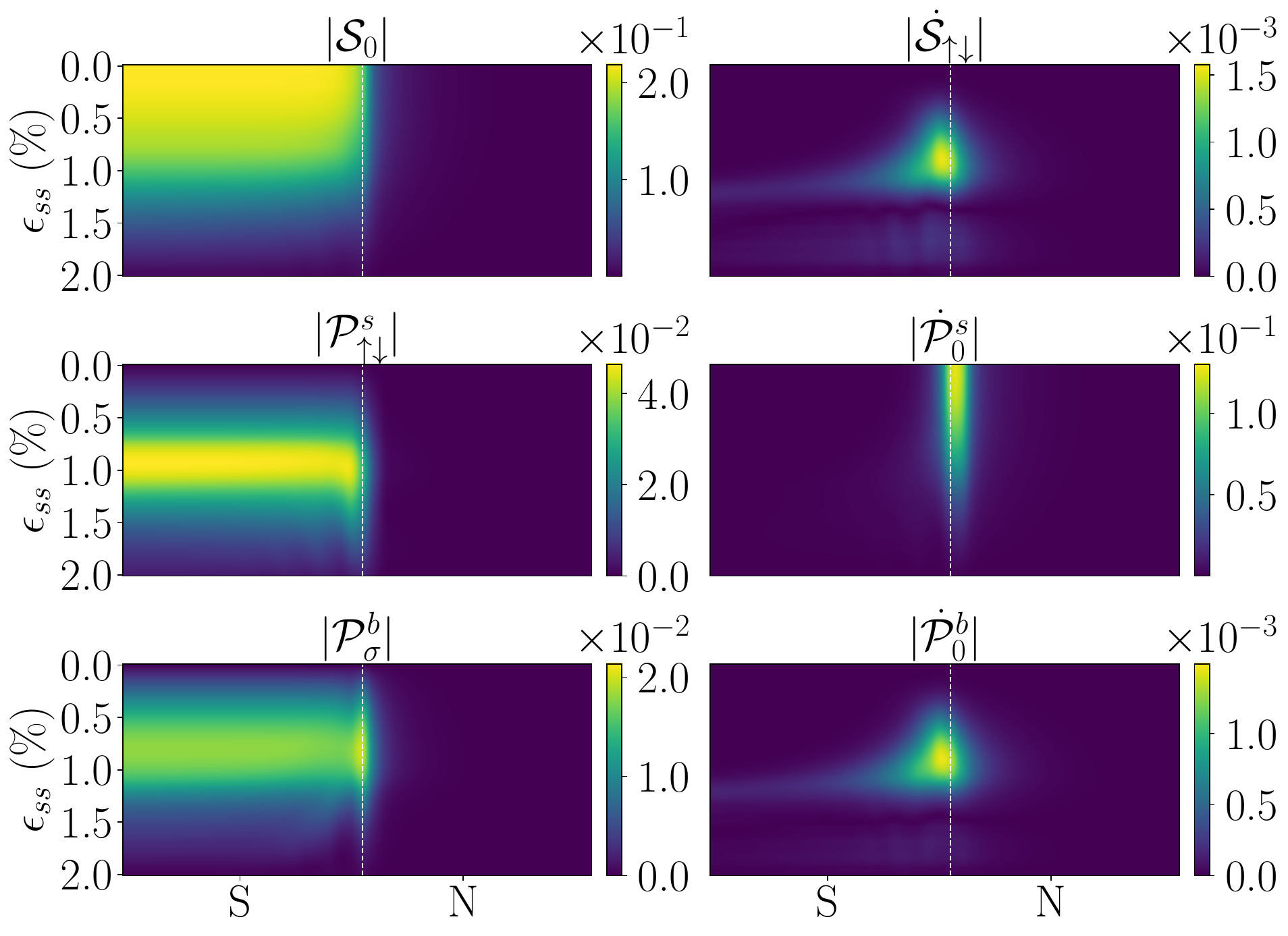}
    \caption{The pairing amplitudes around the SN interface for increasing strain in the superconductor. The first column shows the non-zero even-frequency pairings and the second column is the relative time derivative of the odd-frequency pairings at $\tau=0$. The spin-polarized p-wave pairings are equal in magnitude $|\mathcal{P}^b_\sigma| = |\mathcal{P}^b_{-\sigma}|$. The derivatives are compared to the amplitudes of the corresponding odd-frequency pairings in Fig. S.1 in the Supplementary Information. System parameters are given in the main text.}
    \label{fig:Magnitudes_interface}
\end{figure}

In \cref{fig:Magnitudes_interface}, we show the even-frequency pairing amplitudes and the relative time derivatives of the finite odd-frequency pairings at the interface. The even-frequency amplitudes originate from the bending-strain via the spin-asymmetry introduced by the effective spin-orbit coupling in the superconductor, while the interface between the materials induces the odd-frequency pairings. The even-frequency p-wave triplets $\mathcal{P}^s_{\su\sd}$ persist throughout the bulk, and these triplets have an interface-dependence on both Andreev and normal reflections \cite{Cayao2018InterfaceRashba}. They vanish for $\alpha_N = 0$, in contrast to $\mathcal{P}^s_0(\tau)$, which remains finite, as can be seen in \cref{fig:Magnitudes_interface}. Furthermore, since the superconductor is strained and the normal metal is not, the curvature-induced spin-orbit coupling vanishes at the interface. This interfacial change acts on spins like a local spin splitting in the binormal direction, giving rise to an odd-frequency $\mathcal{S}_{\su\sd}(\tau)$-wave triplet and $\mathcal{P}^b_0(\tau)$-wave singlet contribution. 

By considering the mechanisms for interconversion between the different symmetries of the superconducting correlations, we can therefore detangle the independent contributions generated by the interface and the strain. By examining \cref{fig:Magnitudes_interface}, we can see how their respective magnitudes can be tuned with the strain $\epsilon_{ss}$, for which a number of external control mechanisms exist, for example mechanical or thermoelectric manipulation, voltage-control from piezoelectric actuators \cite{Salamone2024spin_valves}, or light-control via a photostrictive substrate \cite{Kundys2015}. 

For completeness, we compare the derivatives in \cref{fig:Magnitudes_interface} with the magnitudes of the corresponding odd-frequency pairings in Fig. S.1 in the Supplementary Information. There we see that the amplitudes -- which remain within the same order for relative times within that order -- are comparable to the derivative evaluated at relative time $\tau=0$. 

\subsection{SNS junction with bending-strain in S}\label{Subsec:Results SNS}
Lastly, we consider an SNS junction with strained superconductors and an un-strained, uncurved normal metal, as illustrated in  \cref{fig:curvilinear coordinates}. A charge-current flows across the normal metal when there is a phase difference between the superconductors \cite{Josephson1962,Tinkham}. In addition to the charge current, we find that a spin-current flows across the junction whenever there is bending-strain in the superconductors. Spin current is in general not a conserved quantity since SOC prevents spin from being a good quantum number \cite{bulou2021magnetism}. However, in our model there is no strain or SOC within the normal metal, and so any spin-current is therefore conserved in this region. 

\begin{figure}[hbt]
    \centering
    \includegraphics[width=0.48\textwidth]{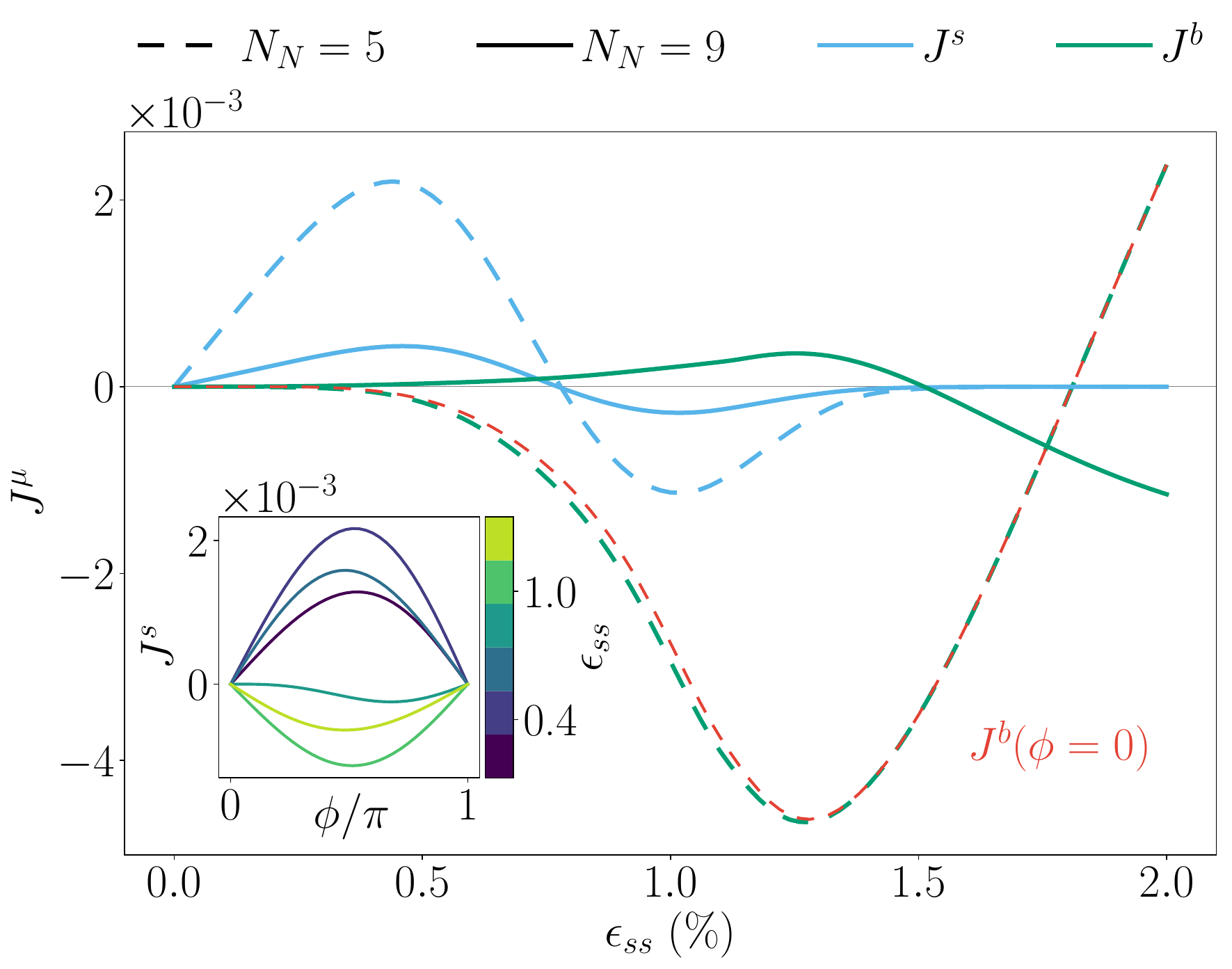}
    \caption{The nonzero spin-current components at the center of the normal metal for two different lengths $N_N=(5,9)$ at phase difference $\phi=\pi/2$. We have used $N_S=50$ sites in the superconductors and $N_b = 2N_S +N_N$ in both cases. The remaining parameters are identical to the main text. $J^b$ for $N_N=5$ at zero phase difference $\phi=0$ is shown in red. The inset shows the sinusoidal form of $J^s$, with a change in sign as a function of strain.}
    \label{fig:spin_current_components} 
\end{figure}

In \cref{fig:spin_current_components}, we show the tangential and binormal components of the spin current $\bm{J}$ at the center of the normal metal. The sign of $J^b$ depends on the length of the normal metal and strain, resulting in so-called $0-\pi$ oscillations as a function of both length and the effective spin-orbit coupling strength experienced, shown here as a function of the bending-strain in the superconductor. The $0-\pi$ oscillations in $J^b$ as a function of length are included in Fig. S.2, in the Supplementary Information for completeness. Changes in current-direction as indicated by $0-\pi$ oscillations are well known in Josephson junctions with a ferromagnetic weak link \cite{Ryazanov2001, Kontos2002, Birge2024}, but in SNS junctions they are so-far only found when there exists a non-equilibrium electron distribution in N \cite{Baselmans1999}. This result therefore indicates that, with regards to generating a Josephson current, strain, or effective SOC, acts on the normal metal in a similar way to a Zeeman interaction in a ferromagnet. A subgap spin Josephson effect has also been suggested in the topologically nontrivial phase of topological Josephson junctions with a magnetic field and SOC in the superconductors \cite{Mao2022}, and a superconducting diode effect has been noted due to a combination of SOC and magnetic field or intrinsically p-wave superconductors \cite{mao2024universal}. In our case, we show that this current can appear when using conventional s-wave superconductors with bending-strain and a simple normal metal weak link. 

At low strain, we see in \cref{fig:spin_current_components} that the dominant Josephson-current is carried by spins polarized in the tangential direction, while $J^b$ takes precedence after the $0-\pi$ transition of $J^s$, due to the increasing spin-orbit coupling. The sinusoidal current-phase relation of $J^s$ is shown in the inset of \cref{fig:spin_current_components}. The binormal component of the spin current consists of a constant value plus a cosine term. Therefore, it is possible to create a small, \textit{pure} spin-current $\bm{J} = J^{b}\: \hat{\mathcal{B}}(s)$ flowing through the junction. It has been shown that the proportion of spin current in ferromagnetic Josephson junctions can be tuned by the phase difference \cite{Jacobsen2016}. Here, it is remarkable to note both that a spin current can be \textit{induced solely} from applied strain in a conventional SNS Josephson-junction, and further that the direction of this current can be \textit{tuned} by the magnitude of this applied strain.

\begin{figure}[hbt]
    \centering
    \includegraphics[width=0.48\textwidth]{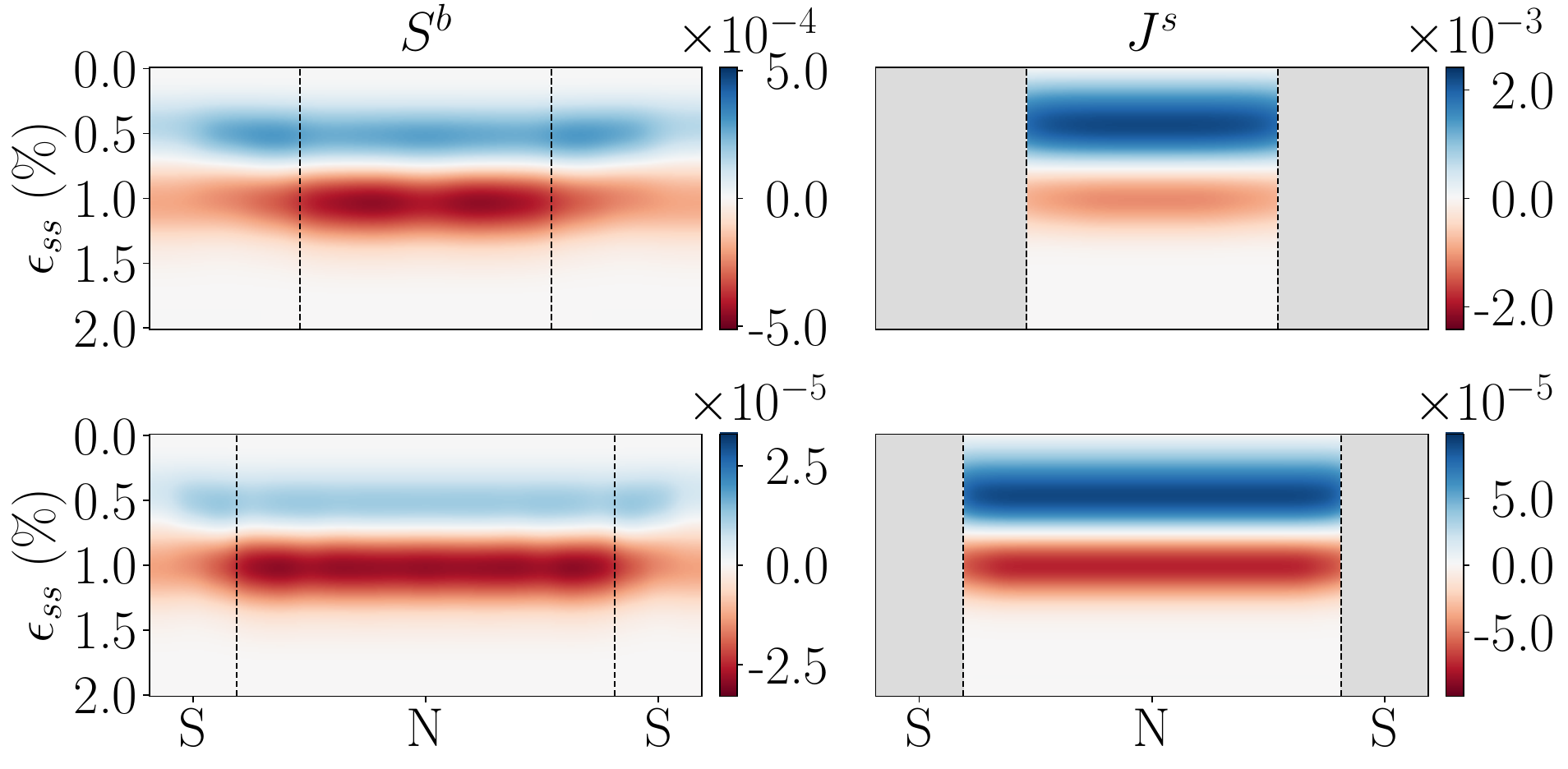}
    \caption{The spin-magnetization $S^b$ and spin-current component $J^s$ within the normal metal, for phase difference $\phi=\pi/2$ between the superconductors. The first row has $N_N = 5$ and the second row $N_N = 13$. We have used $N_S=50$ and $N_b = N_s$, while the rest of the parameters are identical to the main text.}
    \label{fig:Ps_wave and mag Sb}
\end{figure}

In \cref{fig:Ps_wave and mag Sb}, we show the spin-current $J^s$, this time as a function of both strain and position in $N$, alongside the induced spin-magnetization $S^b$. A combination of SOC and charge current in normal metals generically leads to spin accumulation \cite{Manchon2015}; the charge current shifts the Fermi contour along the $k_s$-axis in momentum space, which, in the presence of SOC, creates a finite spin magnetization in the binormal direction. The appearance of a spin-magnetization can therefore be attributed to the Edelstein effect \cite{Edelstein1990original}. In our numerical scheme, we split the phase difference equally between the two superconductors by fixing the phases in the superconductors. Therefore, the spin accumulation is strongest within the normal metal, as seen in \cref{fig:Ps_wave and mag Sb}.  The diagonalization gives proximity-induced pairings in the normal metal, with a phase gradient $\partial_s \phi$. This gradient diminishes away from the interfaces within the superconductors, so the spin magnetization [$S^b$] and charge current will eventually decay there.

We see from \cref{fig:Ps_wave and mag Sb} that the switch in magnetization can be determined from the $0-\pi$ switching profile of $J^s$. The principal reason why no magnetic field or intrinsic p-wave order is required in this case is due to the combination of effective-SOC and interfacial effects on rotating the proximity-induced correlations. It is inherently linked to the thin-film nature of the system, which retains an additional term in the cross-product of the spin vector and the nearest-neighbour vector [\cref{eq:BdG_hamiltonian}] compared with the one-dimensional analysis of most triplet transport analysis to date. When considering two dimensions, we get a $\sigma_T \hat{k}_s$-term in the $sb$-plane in the presence of a phase gradient. This creates spin-polarized triplets that can engender a magnetization in the binormal direction, and plays the role typically provided by a magnetic field. We therefore see that strain can be an effective generator and controller/switcher of spin-current and magnetization in SNS junctions. 

\section{Discussion}\label{Sec:Disc}

We have considered the effect of bending-strain in thin films of clean, conventional, s-wave superconductors, and its proximity-induced effects on the symmetries of the superconducting correlations persisting in SN and SNS heterostructures. In the strained superconductor alone, we find that the strain reduces the s-wave gap while increasing even-frequency p-wave triplets both with and without spin polarization throughout the superconductor. Including a normal-metal interface leads to the additional generation of odd-frequency pairings -- odd-frequency $\mathcal{P}^s_0(\tau)$ from the interface alone, and a contribution to $\mathcal{S}_{\su\sd}(\tau)$-wave triplet and $\mathcal{P}^b_0(\tau)$-wave singlet from both the interface and strain -- and their magnitudes can be tuned by the strain. Using these principles, we find that the strain alone can induce a spin-current in SNS junctions, and tuning the strain can switch the current direction. The presence of a spin-polarized current results in a strain-switchable magnetization in the normal metal, and we show that this pure spin current can be generated in the junction without a magnetic field, and even when the phase difference in the junction is $\phi=(0,\pi)$. Considering the many control mechanisms for nanomechanical strain, this may find several uses in superconducting spintronics and quantum logic circuits, from control of superconducting qubits to controllable memories \cite{Birge2024}.

We showed in \cref{Subsec:Results S} that bending-strain in an isolated superconducting film induces p-wave triplet pairings throughout the superconductor for uniform curvature. Their magnitude is always around one order lower than the s-wave singlet, and die out as the parent correlation diminishes. Since the superconducting order parameter started out fully s-wave singlet, our self-consistency calculations included only the s-wave gap. We expect that if the p-wave correlations are included in the self-consistency requirement, p-wave pairings may stabilize to form the majority pairing population for increasingly high strain. For non-constant curvatures in the film, the gap would reduce locally in areas with high strain, and the p-wave contribution would increase only locally. We would also expect the sign of the curvature in each arm of the SNS junction to affect the ground state \cite{skarpeid2024non}. For an $S$-shaped junction with opposite but equal curvatures in each arm, we would expect the spin magnetization to become vanishingly small as the spin polarization contributions from each interface would cancel \cite{skarpeid2024non}.

In this article we have shown the effects of bending-strain magnitudes of up to $2\%$, with all the major effects we report generally achievable with $\epsilon_{ss}< 1\%$. We have shown how thicker materials can support superconductivity at higher strains, but since materials typically become very brittle at cryogenic temperatures, it is prudent to limit strain to around $1\%$. In addition, while a number of techniques are being developed for designing curvilinear superconductors in three dimensions \cite{fomin2022perspective}, most devices for dynamically imparting strain in-situ, such as piezoelectric actuators \cite{Salamone2024spin_valves,Trotta2012} are currently restricted to this same order. 

Bending-strain translates into an effective SOC with symmetry breaking axis in the radial direction. Since the arc length defining the radius of curvature changes for every layer in the film, the total bending-strain of any one sample depends on the film thickness according to \cref{eq:strain}. We showed results for thicknesses $N_n=(3,6,9)$, and for equal strains these thicknesses have both a different curvature and a different corresponding effective SOC. We provide an overview of their comparative values in table S.1., in the Supplementary Information, were we for convenience also compare the SOC energy with respect to the superconducting gap. When combining the radial bending-strain with interfacial-Rashba SOC in thin-film heterostructures, these two different symmetry-breaking axes combine to give a nuanced mixture of odd- and even-frequency correlations. They can be harnessed in conjunction with the spontaneous charge current that exists in an SNS Josephson junction, giving pure spin-current even at zero phase difference in the junction. The spin current can drive a switchable magnetization in the normal metal. Strain alone may therefore also induce a superconducting diode effect \cite{mao2024universal}.

In our approach, we have assumed that the imparted strain is sufficiently small to not significantly affect the electron-phonon interaction. However, to the best of our knowledge there is as yet no clear experimental data indicating for what strains this assumption breaks down. For larger strains, we would expect a significant modification of the electron bands due to lattice deformation, and a shift in the electron-phonon interaction between compressive and tensile regions. It would be instructive to further explore the nature of this change in the interaction in highly strained superconductors in the future.

In further development of this setup, it would be possible to include intrinsic Dresselhaus SOC in the normal metal, as well as out-of-plane symmetry breaking Rashba SOC induced via a substrate. Furthermore, since the normal metal undergoes no in-situ deformation, it would also be possible to include extrinsic Rashba SOC from gating, giving an additional control over the interconversion mechanisms for the superconducting correlations. While we also restricted the analysis in this study to conventional superconductors, a natural extension would be to consider unconventional materials. For chiral p-wave superconductors for example, bending-strain may split the Fermi surface into two helicity bands \cite{huang2024generic}, potentially leading to the emergence of chiral Majorana edge states \cite{uchihashi2021surface,Gentile2015Edge_states,ahmed2024}. For unconventional or multiband superconductors, modifications to the electron-phonon interaction may play an even bigger role, and strain-deformation of the bands in two-band superconductors could have profound implications for stabilizing superconductivity at high magnetic fields \cite{Salamone2023}.

\section{Methods}\label{Sec:Methods}
This section summarizes the theoretical and numerical framework used to derive the results presented above. We introduce geometric curvature in superconductors using the Frenet-Serret frame in \cref{SubSec:Methods Frenet-Serret}, and define the continuum Hamiltonian with curvature-induced bending-strain in \cref{SubSec:Methods Strain SOC}. We discretize the Hamiltonian to give the tight-binding model in \cref{SubSec:Methods discretization}, and in \cref{SubSec:Methods Numerical} we provide details of the numerical algorithm and material parameters employed to produce the examples that are analyzed in the main text.

\subsection{Frenet-Serret frame}\label{SubSec:Methods Frenet-Serret}
We parametrize a curved superconductor using $\bm{R}(s,n,b) = \bm{r}(s) + n \hat{\mathcal{N}}(s) + b \hat{\mathcal{B}}(s)$, where $\bm{r}(s)$ characterizes the curve along the arclength $s$, and the parameters $n$, and $b$ are the normal and binormal coordinates, respectively. The orthonormal basis vectors for the parametrization are the tangential $\hat{\mathcal{T}}$, normal $\hat{\mathcal{N}}$ and binormal $\hat{\mathcal{B}}$ directions, as illustrated in \cref{fig:curvilinear coordinates}.

The Frenet-Serret formulas, which describe the change of the curvilinear unit vectors $\hat{e}_\mu$,  can be summarized as
\begin{align}\label{Eqn:FS}
    \begin{pmatrix}
        \partial_s\: \hat{\mathcal{T}}(s)\\
        \partial_s\: \hat{\mathcal{N}}(s)\\
        \partial_s\: \hat{\mathcal{B}}(s)
    \end{pmatrix} = \begin{pmatrix}
        0 & \kappa(s) & 0\\
        -\kappa(s) & 0 & 0\\
        0 & 0 & 0
    \end{pmatrix} \begin{pmatrix}
        \hat{\mathcal{T}}(s)\\
        \hat{\mathcal{N}}(s)\\
        \hat{\mathcal{B}}(s)
    \end{pmatrix} .
\end{align}
Here we have introduced the geometric curvature function $\kappa(s)$, which for a uniform curve is simply the inverse of the radius of the curve (see \cref{fig:curvilinear coordinates}). Using \cref{Eqn:FS}, we can write the basis vectors as derivatives of $\bm{R}(s,n,b)$. We express them as $\bm{e}_\mu = \partial_\mu \bm{R} = h_\mu \hat{e}_\mu$, where $h_\mu$ are scale factors. The metric tensor is defined as ${\mathcal{G}_{\lambda\mu} = \bm{e}_\lambda \cdot \bm{e}_\mu}$,  and can be written as
\begin{align}
    \mathcal{G}_{\lambda\mu} = \begin{pmatrix}
        h_s^2 & 0 & 0\\
        0 & 1 & 0\\
        0 & 0 & 1
    \end{pmatrix},
\end{align}
where $h_s = 1 - n\kappa(s)$. It lowers contravariant indices, while its inverse, denoted ${\mathcal{G}^{\lambda\mu}}$, raises covariant indices.

For a circular geometry, the curvature is constant $\kappa(s) = \kappa$, and the parametrization can be written as
\begin{align} 
    \bm{r}(s) = {-}\frac{1}{\kappa} \cos(\kappa s)\: \hat{\bm{e}}_x +  \frac{1}{\kappa} \sin(\kappa s)\: \hat{\bm{e}}_y. \label{eq:r_circ}
\end{align}
The orthogonal curvilinear unit vectors become $\hat{\mathcal{T}}(s) = \partial_s \bm{r}(s) $, $ \hat{\mathcal{N}}(s) = {-} \partial_s \hat{\mathcal{T}}(s) / |\partial_s \hat{\mathcal{T}}(s)| $ and $ \hat{\mathcal{B}}(s) = \hat{\mathcal{T}}(s) \times \hat{\mathcal{N}}(s) $. For the circular parametrization [\cref{eq:r_circ}] this gives
\begin{align} \label{eq:parametrization}
\begin{split}
    \hat{\mathcal{T}}(s) &= \sin(\kappa s)\: \hat{\bm{e}}_x + \cos(\kappa s)\: \hat{\bm{e}}_y,\\
    \hat{\mathcal{N}}(s) &= -\cos(\kappa s)\: \hat{\bm{e}}_x + \sin(\kappa s)\: \hat{\bm{e}}_y,\\
    \hat{\mathcal{B}}(s) &=  \hat{\bm{e}}_z.
\end{split}
\end{align}

\subsection{Continuum Hamiltonian and strain-induced spin-orbit coupling}\label{SubSec:Methods Strain SOC}
The covariant Hamiltonian for describing the motion of electrons in the presence of spin-orbit coupling is 
\begin{align} \label{eq:Hamiltonian}
    \mathcal{H} = -\frac{\hbar^2 \mathcal{G}^{\lambda\mu}}{2m^*} \mathcal{D}_\lambda \mathcal{D}_\mu + \frac{i\hbar}{m^*} \frac{\epsilon^{\lambda\mu\nu}}{\sqrt{G}} \alpha_\lambda \sigma_\mu \mathcal{D}_\nu,
\end{align}
where $G$ is the determinant of the metric tensor, $\hbar$ is the reduced Planck constant, $m^*$ is the effective electron mass, $\alpha_\lambda$ are components of the spin-orbit vector, $\sigma_\mu$ the Pauli vector components, and $\epsilon^{\lambda\mu\nu}$ is the Levi-Civita symbol. We define the covariant spin-orbit field as $\mathcal{A}^\nu = \epsilon^{\lambda\mu\nu} \alpha_\lambda \sigma_\mu / \hbar \sqrt{G}$, which allows us to rewrite the Hamiltonian in \cref{eq:Hamiltonian} to show a local SU(2) gauge invariance \cite{Bergeret2013,Bergeret2014,Jacobsen2015b}
\begin{align}
    \mathcal{H} = -\frac{\hbar^2 \mathcal{G}^{\lambda\mu}}{2m^*} \big(\mathcal{D}_\lambda - i \mathcal{A}_\lambda \big) \big(\mathcal{D}_\mu - i \mathcal{A}_\mu \big).
\end{align}
The covariant derivative $\mathcal{D}_\lambda$ of a covariant vector $v_\mu$ is defined as \cite{Pressley2010}
\begin{align}
    \mathcal{D}_\lambda v_\mu = \partial_\lambda v_\mu - \Gamma^\nu_{\lambda\mu} v_\nu.
\end{align}
 The Christoffel symbols are related to the metric tensor, and are defined as \cite{Pressley2010}
\begin{align}
    \Gamma^\nu_{\lambda\mu} = \frac{1}{2} \mathcal{G}^{\nu\nu}\big( \partial_\mu \mathcal{G}_{\nu\lambda} + \partial_\lambda \mathcal{G}_{\nu\mu} - \partial_\nu \mathcal{G}_{\lambda\mu}\big) .
\end{align}

When curvature is the result of lattice deformation, the associated strain can be defined as the difference in arc length between the center $L(0)$ and the outermost layer $L(n)$ normalized by $L(0)$ (see \cref{fig:curvilinear coordinates}). The arclength is given by $L(n) = \theta(R+n)$, where $\theta$ is the subtending angle. Thus, we can write the strain associated with the geometric deformation as \cite{Ortix2011,gentile2013strain}
\begin{align} \label{eq:strain}
    \epsilon_{ss} = \frac{\theta (R + n) - \theta R}{\theta R}=\kappa(s) n.
\end{align}
This in turn gives rise to an additional potential in the material \cite{Bardeen1950,Walle1989} and, therefore, an electric field $\bm{E} \sim \kappa(s) \hat{\mathcal{N}}(s)$ for small strain. An electron moving through this electric field will experience an effective magnetic field $\bm{B} \sim \bm{p} \times \bm{E}$ through the Zeeman interaction in its rest frame. Since the electric field is oriented in the normal direction, it allows us to express the resulting spin-orbit vector as $\bm{\alpha} = \alpha_N \hat{\mathcal{N}}(s)$, where the coefficient $\alpha_N$ quantifies the strain-induced spin-orbit coupling, being proportional to the local curvature, given by $\alpha_N(s) = a_N \kappa(s)$, with proportionality constant $a_N$. Thus, the continuum Hamiltonian for a thin film around $n=0$, can be written as 
\begin{align}\label{eq:cont_hamiltonian}
    \begin{split}
        \mathcal{H} = &-\frac{\hbar^2}{2m^*} \nabla^2 -\frac{i \hbar^2}{m^*} \sigma_T(s)\: \alpha_N(s)\: \partial_b\\ &+\frac{i\hbar^2}{2m^*} \sigma_B \bigl[\alpha_N(s)\: \partial_s + \partial_s\: \alpha_N(s) \bigr]\:,
    \end{split}
\end{align}
where $\sigma_T(s) = - \sin\left[\kappa(s) s\right] \sigma_x + \cos\left[\kappa(s) s\right] \sigma_y$ and $\sigma_B = \sigma_z$, which follows directly from \cref{eq:parametrization}.

\subsection{Discretization and diagonalization}\label{SubSec:Methods discretization}
Discretizing the continuum Hamiltonian [\cref{eq:cont_hamiltonian}] gives the tight-binding Bogoliubov–de Gennes Hamiltonian:
\begin{align}\label{eq:BdG_hamiltonian}
    \mathcal{H} = &-\! \sum_{\langle \bm{i}, \bm{j} \rangle, \sigma} t_{\bm{ij}} c^\dag_{\bm{i}, \sigma} c_{\bm{j}, \sigma} - \sum_{\bm{i}, \sigma} \mu_{\bm{i}} c^\dag_{\bm{i}, \sigma} c_{\bm{i}, \sigma} \nonumber\\
    &- \sum_{\bm{i}} U_{\bm{i}} c^\dag_{\bm{i},\uparrow} c_{\bm{i},\uparrow} c^\dag_{\bm{i},\downarrow} c_{\bm{i},\downarrow} \\
    &-\frac{i}{2} \sum_{\langle \bm{i},\bm{j} \rangle, \sigma, \sigma'} \alpha_N\: \hat{\mathcal{N}}(s) \cdot \left(\bm{\sigma}(s) \times \bm{d_{ij}} \right) {c^\dag_{\bm{i},\sigma} c_{\bm{j},\sigma'},} \nonumber
\end{align}
where $t$ is the hopping amplitude, $\mu_{\bm{i}}$ is the chemical potential, $U_{\bm{i}}$ is the local on-site attraction that gives rise to superconductivity. The nearest neighbor vector $\bm{d_{ij}}$ points from lattice coordinates ${\bm{i}=(i_s, i_b)}$ to ${\bm{j}=(j_s, j_b)}$, and $\bm{d_{ij}} = - \bm{d_{ji}}$. 

We assume periodic boundary conditions in the binormal direction. Therefore, we Fourier transform the electron creation and annihilation operators,
\begin{align}
\begin{split}
    c_{i_s, k_b, \sigma} = \frac{1}{\sqrt{N_b}} \sum_{i_b} c_{i_s, i_b, \sigma} e^{i k_b i_b} , \\
    c^\dag_{i_s, k_b, \sigma} = \frac{1}{\sqrt{N_b}} \sum_{i_b} c^\dag_{i_s, i_b, \sigma} e^{-i k_b i_b},
\end{split}
\end{align}
where $N_b$ represents the total number of lattice sites in the binormal direction. Since the binormal lattice index does not appear in any of the following expressions, we denote the tangential $i_s \equiv i$ and the binormal momentum $k_b \equiv k$. We choose the Fourier-transformed Nambu$\otimes$spin space basis
\begin{align} 
\begin{split}
    \hat{B}^\dag_{i, k} = \begin{pmatrix}
        c^\dag_{i, k, \su} & c^\dag_{i, k, \sd} & c_{i, -k, \su} & c_{i, -k, \sd}
    \end{pmatrix},\\
    W^\dag_{k} = \begin{pmatrix}
        \hat{B}^\dag_{1, k} & \hat{B}^\dag_{2, k} & \hat{B}^\dag_{3, k} & \dots & \hat{B}^\dag_{N_s, k}  
    \end{pmatrix},
\end{split}
\end{align}
which we use to write the Hamiltonian as
\begin{align} \label{eq:k_sum_nambu_spin_space}
    \mathcal{H} = -\frac{1}{2} \sum_{i, j, k} \hat{B}^\dag_{i, k} \hat{H}_{i, j, k} \hat{B}_{j, k}  = -\frac{1}{2} \sum_{k} W^\dag_{k} H_{k} W_{k},
\end{align}
where we have neglected constant shifts to the energy. The site-dependent matrix in the above sum can be written as
\begin{align} \label{eq:discretized_hamiltonian}
\begin{split}
    \hat{H}_{i, j, k} =&\: \big( 2t \cos ka + \mu_{i}\big) \delta_{i, j}\: {\tau}_3 \otimes {\sigma}_0 \\
    & + t\: \big(\delta_{i, j + 1} + \delta_{i, j - 1} \big)\: {\tau}_3 \otimes {\sigma}_0 \\
    &+ \delta_{i, j}\: \big(i\Delta_{i}\:  {\tau}^+ - i\Delta_{i}^*\:  {\tau}^- \big) \otimes {\sigma}_y  \\
    &+\alpha_{ii}\: \delta_{i, j}\: \sin ka\: \hat{\sigma}_T \\
    & + \frac{i}{2} \alpha_{ij} \big(\delta_{i, j+1} - \delta_{i, j -1} \big)\: {\tau}_0 \otimes {\sigma}_{B},
\end{split}
\end{align}
where $\alpha_{ij}=\alpha_N(s_i) + \alpha_N(s_j)$, and $\tau^\pm = \tau_1 \pm i\tau_2$, where the $\tau$ matrices are Pauli matrices that operate in Nambu space. Furthermore, the matrix $\hat{\sigma}_{T,i} = \diag(\sigma_{T,i}, \sigma^{*}_{T,i})$, where $\sigma_T(s) = \bm{\sigma} \cdot \hat{\mathcal{T}}(s)$, and $a$ is the lattice constant.\\

To diagonalize the Hamiltonian and obtain the eigenenergies we consider the eigenvalue problem $H_{k} \Phi_{n, k} = \mathcal{E}_{n, k} \Phi_{n, k}$, which has $4N_s$ eigenvalues, where $N_s$ is the total number of lattice sites in the tangential direction. It is equivalent to writing
\begin{align}\label{eq:eigenproblem}
    \sum_{j} \hat{H}_{i, j, k} \hat{\phi}_{j, n, k} = \mathcal{E}_{n, k} \hat{\phi}_{i, n, k},
\end{align}
where 
\begin{align} \label{eq:eigenvectors}
\begin{split}
    \hat{\phi}^\dag_{i, n, k} &= \big(u^*_{i, n, k},\: v^*_{i, n, k},\: w^*_{i, n, k},\: x^*_{i, n, k} \big),\\
    {\Phi}_{n, k} &= \big(\hat{\phi}_{1, n, k},\: \hat{\phi}_{2, n, k},\: \ldots,\: \hat{\phi}_{N_s, n, k} \big) .
\end{split}
\end{align}
Diagonalizing the $4N_s\times4N_s$ matrix gives ${H_{k} = P_{k} D_{k} P^\dag_{k}}$, where ${P_{k} = [{\Phi}_{n, k}]}$ is a matrix containing the eigenvalues of $H_k$. The matrix ${D_{k}}$ contains the eigenvalue $\mathcal{E}_{n, k}$ on the n'th diagonal. Therefore, the total Hamiltonian becomes
\begin{align}
    \mathcal{H} = - \frac{1}{2} \sum_{n, k} \mathcal{E}_{n, k} \gamma^\dag_{n, k} \gamma_{n, k},
\end{align}
where $\gamma_{n, k}$ is the n'th element of $\Gamma_{k} = P^\dag_{k} W_{k}$.

There are pairs of linearly dependent quasiparticle operators ${\gamma_{n, k} = \gamma^\dag_{n, -k}}$ that give pairs of identical eigenvalues ${\mathcal{E}_{n,k} = - \mathcal{E}_{n,-k}}$ with opposite sign. The momentum index lies in the first Brillouin zone ${ka \in [-\pi,\pi)}$, and for periodic boundary conditions, the values are restricted by $ka = 2\pi m / N_b$, where $m$ is an integer. Thus, we split the momentum sum into three contributions to obtain a Hamiltonian with only linearly independent quasiparticle operators \cite{gonzalez2021superconductivity}:
\begin{align}
\begin{split}
    \mathcal{H} =\: & -\!\!\! \sum_{n, 0 < ka < \pi } \mathcal{E}_{n, k} \gamma^\dag_{n, k} \gamma_{n, k} \\
    &- \sum_{\mathcal{E}_n \geq 0} \big( \mathcal{E}_{n, 0} \gamma^\dag_{n, 0} \gamma_{n, 0} + \mathcal{E}_{n, \pi} \gamma^\dag_{n, \pi} \gamma_{n, \pi} \big).
\end{split}
\end{align}
For zero momentum and $ka=\pi$ we used the following relations for the pairs of energy eigenvalues: ${\mathcal{E}_{n,0} = - \mathcal{E}_{n+2N_s, 0}}$ and ${\mathcal{E}_{n,\pi} = - \mathcal{E}_{n+2N_s, \pi}}$. All expectation values are expressed similarly, so we define a new sum
\begin{align} \label{eq:new_sum}
    \displaystyle \sum_{n, k}^{'} = \sum_{\mathcal{E}_n \geq 0, k = 0} + \sum_{n, 0 < ka < \pi } + \sum_{\mathcal{E}_n \geq 0, ka = \pi} \:.
\end{align}
The old fermion operators can now be expressed using the new quasiparticle operators:
\begin{align} \label{eq:old operators using uvwx}
\begin{split}
    c_{i, k, \su} &= \sum_n u_{i, n, k} \gamma_{n, k},\:\:\:\:\:\: c_{i, k, \sd} = \sum_n v_{i, n, k} \gamma_{n, k},\\
    c^\dag_{i, -k, \su} &= \sum_n w_{i, n, k} \gamma_{n, k},\:\:\: c^\dag_{i, -k, \sd} = \sum_n x_{i, n, k} \gamma_{n, k},    
\end{split}
\end{align}
which are useful in extracting numerical estimates for observables of interest.

\subsection{Numerical details}\label{SubSec:Methods Numerical}
\begin{figure}[hbt]
    \centering
    \includegraphics[width=0.45\textwidth]{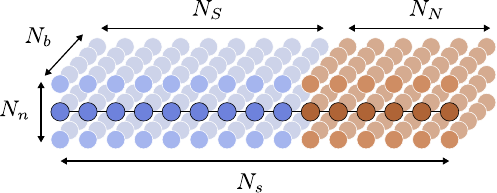}
    \caption{The illustration shows the dimensions used in the numerical treatment of an SN bilayer. Assuming periodic boundary conditions in the binormal direction, the thin-film is modeled as a one-dimensional chain. The thickness ($N_n$) provides an effective SOC, specifically in the direction normal to the strained superconductor.}
    \label{fig:LatticeModel}
\end{figure}

In \cref{fig:LatticeModel}, we show the dimensions used in the numerical scheme: the \textit{total} number of lattice sites in the tangential, normal, and binormal direction are respectively denoted $N_s, N_n$, and $N_b$, while the number of lattice sites in the superconductors and normal metal \textit{individually} in the tangential and normal direction are respectively denoted $N_S$ and $N_N$.

We initialize a superconductor(s) with an s-wave gap on a square lattice, which is solved self-consistently with an iterative diagonalization by updating the gap in the Hamiltonian [\cref{eq:discretized_hamiltonian}] with the expression in \cref{eq:gap_bdg} until the absolute difference between iterations falls below a threshold, here set to $10^{-6}$. We set the number of sites in the binormal and tangential direction to be square, with $ N_b = N_s = 99$, and Fourier transform the binormal direction with periodic boundary conditions. The numerical implementation is simplified by selecting an odd number of lattice sites, since the end of the Brillouin zone does not contribute in this case because the binomial momentum is restricted by $ ka = 2\pi m/N_b $, where $ m $ is an integer [in effect, we can ignore the last term in \cref{eq:new_sum}]. While a specific number is chosen for illustration, the results are not sensitive to the number of lattice sites and remain consistent across different configurations, although a more sophisticated model would be expected to show atomic-scale oscillation of the strength of the superconductivity \cite{Guo2004,EliashbergThinFilm2024Ummarino}.

For the results presented in \cref{Sec:Results}, we normalize energy scales to the hopping parameter $t$, and set the attractive potential parameter to $U=1.5$, the chemical potential to $\mu=1$, and the temperature to $T=0.2$. Furthermore, we normalize the curvature to the superconducting coherence length $\alpha_N = \kappa \xi$, with proportionality constant $a_N=1$. This can be used to relate the effective spin-orbit strength to the strain via $\alpha_N = \epsilon_{ss} \xi / n$, where $n=(N_n-1) a /2$ is the length from the center of the film to the outermost layer (see \cref{fig:curvilinear coordinates}), and $a$ is the lattice spacing. Using this, any specific strain values indicated in \cref{Sec:Results} refer to the strain at the upper/lower plane of the superconductor. For illustrative purposes, we selected parameters corresponding to superconducting niobium, with $\alpha_N = 2.3 \epsilon_{ss}/(N_n - 1)$, where $\epsilon_{ss}$ is the strain in percent. The choice of material is arbitrary, and was chosen only to provide a familiar scale in comparison with purely spin-orbit-coupled systems, and the results that were presented will apply to any conventional thin-film superconductor\cite{Blatt1963,Guo2004,Chiang2004,EliashbergThinFilm2024Ummarino}.

In the limiting case $N_n \to \infty$, the Rashba coefficient $\alpha_N\to 0$, suppressing p-wave pairings and spin-currents. Conversely, a strain-induced electric field $\mathbf{E} \propto \kappa \hat{\mathbf{\mathcal{N}}}$ requires $N_n \geq 2$. In order to emphasize the thickness dependence of pairing and spin responses under strain we considered the range $N_n \in (3,9)$, and set $\xi/a=2.3$ so that $\alpha_N$ falls within the common range for SOC-systems. The thin-film Hamiltonian is solved by summing over one-dimensional chains with a binormal momentum index $k$ and averaging over the first Brillouin zone, as described in \cref{eq:k_sum_nambu_spin_space,eq:new_sum}. Strain and curvature are incorporated as spin-orbit coupling, with the Pauli matrices rotating according to \cref{eq:parametrization}. The coordinate system is oriented such that the normal metal, when present, lies in the $y-z$ plane, as illustrated in \cref{fig:curvilinear coordinates}.
%\marked{
%\section*{Data Availability}
%The code used to produce the results presented can be provided by the corresponding author on a reasonable request.}

\section*{Acknowledgments}
We acknowledge funding via the ``Outstanding Academic Fellows'' programme at NTNU and the Research Council of Norway Grant Nos. 302315 and 262633.

\pagebreak
\begin{center}
\widetext
\textbf{\large Supplementary Information:\\ Bending-strain effects in conventional superconductors and superconducting junctions}\\
{\ }\\%
Kjell S. Heinrich,$^1$ Henning G. Hugdal,$^1$ Morten Amundsen,$^1$ and Sol H. Jacobsen$^1$\\
{\ }\\%
\textit{Center for Quantum Spintronics, Department of Physics, NTNU,\\ Norwegian University of Science and Technology, NO-7491 Trondheim, Norway}

\end{center}
%%%%%%%%%% Merge with supplemental materials %%%%%%%%%%
%%%%%%%%%% Prefix a "S" to all equations, figures, tables and reset the counter %%%%%%%%%%
\setcounter{equation}{0}
\setcounter{figure}{0}
\setcounter{table}{0}
\setcounter{page}{1}
\makeatletter
\renewcommand{\theequation}{S\arabic{equation}}
\renewcommand{\thefigure}{S\arabic{figure}}
\renewcommand{\bibnumfmt}[1]{[S#1]}
\renewcommand{\citenumfont}[1]{S#1}
%%%%%%%%%% Prefix a "S" to all equations, figures, tables and reset the counter %%%%%%%%%%

\twocolumngrid

\section{Pair amplitudes}\label{app:expectation values}
In the main text, adhering to the SPOT symmetry restrictions, we defined the s- and p-wave singlet and triplet amplitudes as
\begin{align} 
    \mathcal{S}_{\bm{i}, 0} &= \frac{1}{2} \big[\mean{c_{\bm{i},\su} c_{\bm{i},\sd}} - \mean{c_{\bm{i},\sd} c_{\bm{i},\su}}  \big],\label{eq:definitions s0}\\
    \mathcal{S}_{\bm{i}, \su\sd}(\tau) &= \frac{1}{2} \big[\mean{c_{\bm{i},\su}(\tau)\: c_{\bm{i},\sd}(0)} + \mean{c_{\bm{i},\sd}(\tau)\: c_{\bm{i},\su}(0)} \big],\\
    \begin{split}
        \mathcal{P}_{\bm{i}, 0}^{\bm{m}}(\tau) &= \sum_{\pm}\!\pm  \frac{1}{2} \big[\mean{c_{\bm{i},\su}(\tau)\: c_{\bm{i}\pm\bm{m},\sd}(0)} \\[-5pt]
        &\qquad\qquad - \mean{c_{\bm{i},\sd}(\tau)\: c_{\bm{i}\pm\bm{m},\su}(0)} \big],
    \end{split} \\
   \mathcal{P}^{\bm{m}}_{\bm{i}, \su\sd} &= \sum_\pm\! \pm \frac{1}{2} \big[\mean{c_{\bm{i},\su} c_{\bm{i}\pm\bm{m},\sd}} + \mean{c_{\bm{i},\sd} c_{\bm{i}\pm\bm{m},\su}}  \big], \label{eq:definitions p_t0}\\
    \mathcal{P}^{\bm{m}}_{\bm{i}, \sigma\sigma} &= \sum_\pm \!\pm \mean{c_{\bm{i},\sigma} c_{\bm{i}\pm\bm{m},\sigma}},\label{eq:definitions p_sigma}
\end{align}
where $\bm{m}$ can be $a\hat{\mathcal{T}}(s)$ or $a\hat{\mathcal{B}}(s)$ -- one lattice spacing in the tangential or binormal direction, respectively. The annihilation and creation operators can be expressed in terms of the Bogoliubov quasiparticle operators
\begin{align}
    c_{i, k, \su} &= \sum_n u_{i, n, k} \gamma_{n, k},\label{eq:old operators using uvwx 1}\\ 
    c_{i, k, \sd} &= \sum_n v_{i, n, k} \gamma_{n, k}, \\
    c^\dag_{i, -k, \su} &= \sum_n w_{i, n, k} \gamma_{n, k},\\
    c^\dag_{i, -k, \sd} &= \sum_n x_{i, n, k} \gamma_{n, k}. \label{eq:old operators using uvwx 4}   
\end{align}

The pairing amplitudes can be found by inserting the operators in Eq.~(\ref{eq:old operators using uvwx 1})-(\ref{eq:old operators using uvwx 4}) into (\ref{eq:definitions s0})-(\ref{eq:definitions p_sigma}). This gives the following expressions for the even- and odd-frequency s- and p-wave singlet and triplet amplitudes, and their derivatives
\begin{widetext}
\begin{eqnarray}
        \mathcal{S}_{0, i} &=& \frac{1}{2 N_b} \sum_{n, k}^{'} \Big[x^*_{i, n, k} u_{i, n, k}-w^*_{i, n, k} v_{i, n, k} \Big]\tanh(\beta \mathcal{E}_{n, k}),\\ %%%%%%%%%%%%%%%%%%%%%%%%%%%%%%%%%%%%%%%%%%%%%%%%%%%%%%%%%%%%%%%%%%%%%%%%%%%%%%%%%%%%%%%%%%%%%%%%%%%%%%%%%%%%%%%%%%%%%%%%%%%%%%%%%%%%%%%%%%
        \mathcal{S}_{i,\su\sd}(\tau) &=& \frac{1}{2N_b} \sum_{n,k}^{'} \bigl(x^*_{i,n,k} u_{i,n,k} + w^*_{i,n,k} v_{i,n,k} \bigr) \bigl(f(2\mathcal{E}_{n,k}) e^{2i\mathcal{E}_{n,k} \tau} + f(-2\mathcal{E}_{n,k}) e^{-2i\mathcal{E}_{n,k} \tau} \bigr), \\ %%%%%%%%%%%%%%%%%%%%%%%%%%%%%%%%%%%%%%%%%%%%%%%%%%%%%%%%%%%%%%%%%%%%%%%%%%%%%%%%%%%%%%%%%%%%%%%%%%%%%%%%%%%%%%%%%%%%%%%%%%%%%%%%%%%%%%%%%%
        \dot{\mathcal{S}}_{i,\su\sd}(\tau=0) &=& \frac{-i}{N_b} \sum_{n,k}^{'} \mathcal{E}_{n,k}  \bigl(x^*_{i,n,k} u_{i,n,k} + w^*_{i,n,k} v_{i,n,k} \bigr) \tanh(\beta \mathcal{E}_{n,k}), \\ %%%%%%%%%%%%%%%%%%%%%%%%%%%%%%%%%%%%%%%%%%%%%%%%%%%%%%%%%%%%%%%%%%%%%%%%%%%%%%%%%%%%%%%%%%%%%%%%%%%%%%%%%%%%%%%%%%%%%%%%%%%%%%%%%%%%%%%%%%
        \mathcal{P}_{i,0}^s(\tau) &=& \frac{1}{2N_b} \sum_{\pm, n, k}^{'} \pm \bigg[\big( x^*_{i\pm1, n, k} u_{i,n,k} - w^*_{i\pm1, n, k} v_{i, n, k} \big) f(-2 \mathcal{E}_{n,k}) e^{-2i\mathcal{E}_{n, k} \tau}\nonumber\\
        &&+ \left(w^*_{i, n, k} v_{i\pm1, n, k} - x^*_{i,n,k} u_{i\pm1, n, k} \right) f(2\mathcal{E}_{n, k}) e^{2i \mathcal{E}_{n, k} \tau}\bigg], \\ %%%%%%%%%%%%%%%%%%%%%%%%%%%%%%%%%%%%%%%%%%%%%%%%%%%%%%%%%%%%%%%%%%%%%%%%%%%%%%%%%%%%%%%%%%%%%%%%%%%%%%%%%%%%%%%%%%%%%%%%%%%%%%%%%%%%%%%%%%
         \dot{\mathcal{P}}_{i, 0}^s(\tau=0)  &=& \frac{-i}{N_b} \sum_{\pm, n, k}^{'} \pm \mathcal{E}_{n,k}  \bigg[\big( x^*_{i\pm1, n, k} u_{i,n,k} - w^*_{i\pm1, n, k} v_{i, n, k} \big) f(-2 \mathcal{E}_{n,k}) \nonumber\\
         &&+ \left(x^*_{i,n,k} u_{i\pm1, n, k} - w^*_{i, n, k} v_{i\pm1, n, k} \right) f(2\mathcal{E}_{n, k}) \bigg], \\ %%%%%%%%%%%%%%%%%%%%%%%%%%%%%%%%%%%%%%%%%%%%%%%%%%%%%%%%%%%%%%%%%%%%%%%%%%%%%%%%%%%%%%%%%%%%%%%%%%%%%%%%%%%%%%%%%%%%%%%%%%%%%%%%%%%%%%%%%%
        \mathcal{P}^b_{i, 0}(\tau) &=& \frac{-i}{N_b} \sum_{n,k}^{'} \sin(ka) \left(x^*_{i,n,k} u_{i,n,k} + w^*_{i,n,k} v_{i,n,k} \right) \left(f(2\mathcal{E}_{n,k}) e^{2i \mathcal{E}_{n,k} \tau} + f(-2\mathcal{E}_{n,k}) e^{-2i \mathcal{E}_{n,k} \tau}  \right), \\ %%%%%%%%%%%%%%%%%%%%%%%%%%%%%%%%%%%%%%%%%%%%%%%%%%%%%%%%%%%%%%%%%%%%%%%%%%%%%%%%%%%%%%%%%%%%%%%%%%%%%%%%%%%%%%%%%%%%%%%%%%%%%%%%%%%%%%%%%%
        \dot{\mathcal{P}}_{i, 0}^b(\tau=0) &=& \frac{-1}{N_b} \sum_{n,k}^{'} \mathcal{E}_{n,k} \sin(ka)  \left(x^*_{i,n,k} u_{i,n,k} + w^*_{i,n,k} v_{i,n,k} \right) \tanh(\beta \mathcal{E}_{n,k}), \\ %%%%%%%%%%%%%%%%%%%%%%%%%%%%%%%%%%%%%%%%%%%%%%%%%%%%%%%%%%%%%%%%%%%%%%%%%%%%%%%%%%%%%%%%%%%%%%%%%%%%%%%%%%%%%%%%%%%%%%%%%%%%%%%%%%%%%%%%%%
         \mathcal{P}_{i,\su\sd}^s  &=& \frac{1}{2 N_b} \sum_{n, k, \pm}^{'} \pm \biggl[\left(x^*_{i\pm1, n, k} u_{i, n, k} + w^*_{i\pm1, n, k} v_{i, n, k} \right) f(-2\mathcal{E}_{n, k})\nonumber\\
         &&+ \left(x^*_{i, n, k} u_{i\pm1, n, k} + w^*_{i, n, k} v_{i\pm1, n, k} \right) f(2\mathcal{E}_{n, k}) \biggr], \\ %%%%%%%%%%%%%%%%%%%%%%%%%%%%%%%%%%%%%%%%%%%%%%%%%%%%%%%%%%%%%%%%%%%%%%%%%%%%%%%%%%%%%%%%%%%%%%%%%%%%%%%%%%%%%%%%%%%%%%%%%%%%%%%%%%%%%%%%%%
        \mathcal{P}_{i, \uparrow\su}^b & =& \frac{- 2 i }{N_b} \sum_{n, k}^{'} w^*_{i, n, k} u_{i, n, k}  \sin(ka) \tanh(\beta \mathcal{E}_{n, k}),\\ %%%%%%%%%%%%%%%%%%%%%%%%%%%%%%%%%%%%%%%%%%%%%%%%%%%%%%%%%%%%%%%%%%%%%%%%%%%%%%%%%%%%%%%%%%%%%%%%%%%%%%%%%%%%%%%%%%%%%%%%%%%%%%%%%%%%%%%%%%
        \mathcal{P}_{i, \downarrow\sd}^b & =& \frac{- 2 i }{N_b} \sum_{n, k}^{'} x^*_{i, n, k} v_{i, n, k}  \sin(ka) \tanh(\beta \mathcal{E}_{n, k}). 
\end{eqnarray}
\end{widetext}
We have used that the time-dependent operators are given by $ c_{\bm{i},\sigma}(\tau) = e^{i\mathcal{H}\tau} c_{\bm{i}, \sigma} e^{-i\mathcal{H}\tau}$, where $\mathcal{H}$ is the Hamiltonian and $\tau$ the relative time coordinate. The expectation value of the number operator in Bogoliubov quasiparticle operators becomes ${\mean{\gamma^\dag_{n,k} \gamma_{n', k}} = f(2 \mathcal{E}_{n, k}) \delta_{n,n'}}$, where $f(2\mathcal{E}_{n, k})$ is the Fermi-Dirac distribution, and $\beta$ is the inverse temperature.

\section{Odd-frequency amplitudes}
\setcounter{figure}{0} 
The odd-frequency order parameters are influenced by the relative time coordinate $\tau$, which is scaled by $\hbar/t$ and may appear arbitrary. We have found that the amplitudes remain within the same order of magnitude for relative times that are within the same order. They are also comparable to the derivative evaluated at zero. In \cref{fig:odd_freq_derivatives}, we have plotted the order parameters at $\tau=1$, along with their derivatives evaluated at $\tau=0$. 

\begin{figure}[ht]
    \centering
    \includegraphics[width=0.48\textwidth]{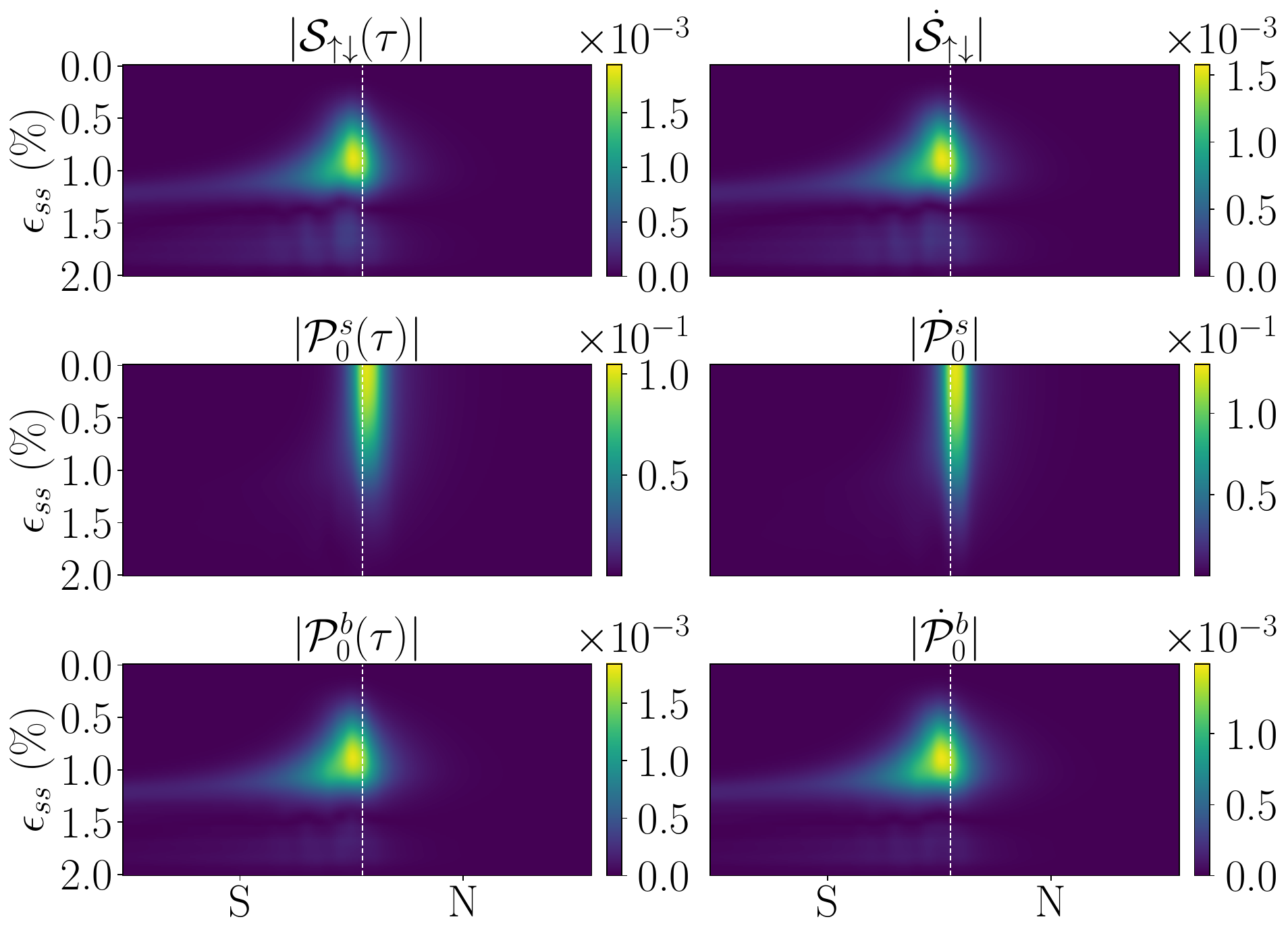}
    \caption{The odd-frequency order parameters that are finite at the superconductor-normal metal interface along with their relative time derivatives. The first column is evaluated at $\tau=1$, and their derivatives in the second column at $\tau=0$.}
    \label{fig:odd_freq_derivatives}
\end{figure}

\section{\texorpdfstring{Spin-current $0-\pi$ oscillations with length}{Spin-current 0-pi oscillations with length}}\label{sec:oscill}
\begin{figure}[hbt]
    \centering
    \includegraphics[width=0.48\textwidth]{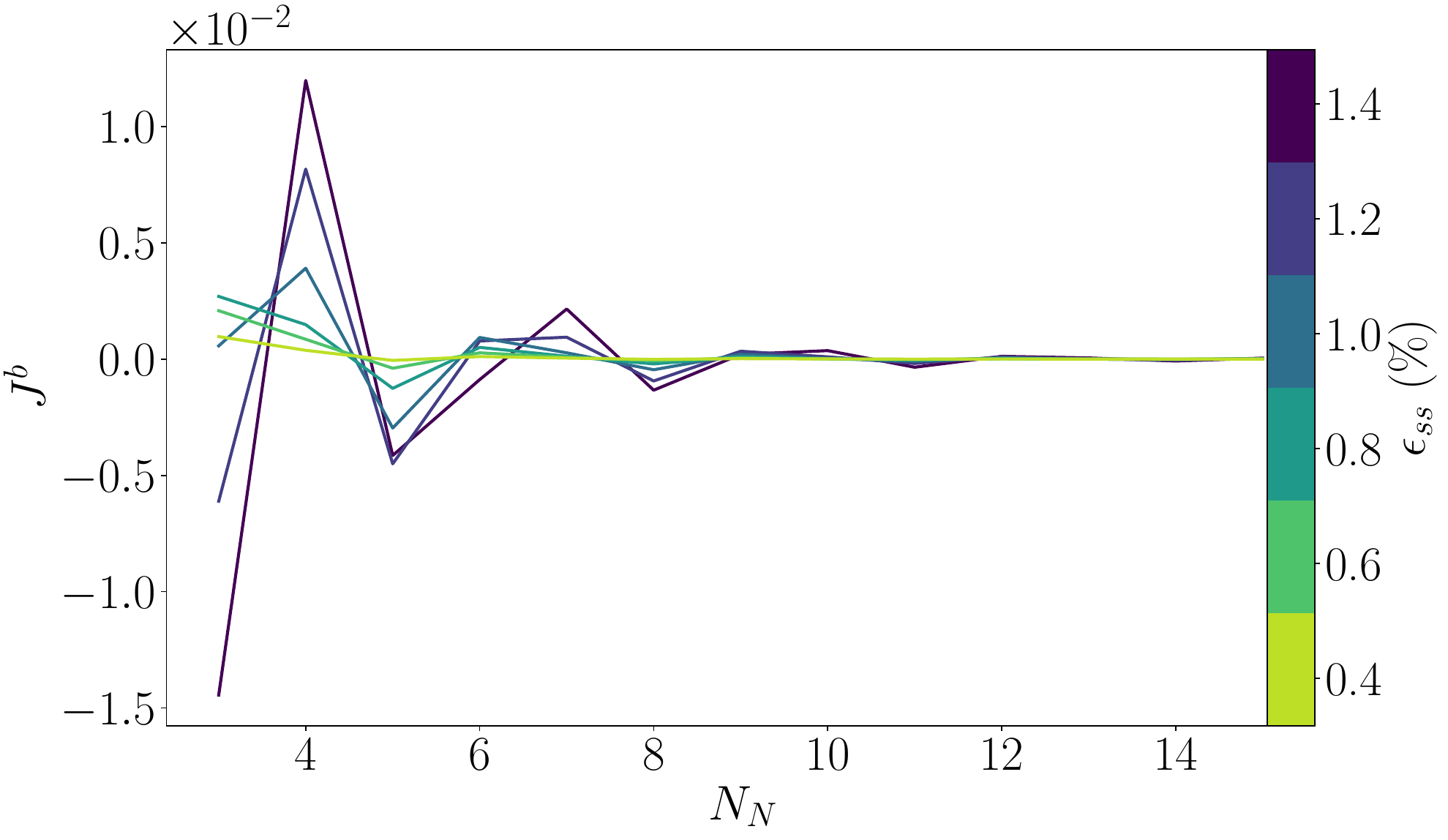}
    \caption{The binormal component of the spin-current exhibits $0-\pi$ oscillations as a function of increasing number of lattice sites in the normal metal, $N_N$. The values are taken from the center of the normal metal. We have used $N_S=40$, $N_b=99$, and $N_n=3$ sites.}
    \label{fig:spin_current_oscillations}
\end{figure}

In an SNS junction, the binormal component of the spin-current in the tangential direction, $J^b$, undergoes $0-\pi$ oscillations as a function of both the applied strain to the superconductors, as seen in Fig. (5) in the main text, and as a function of the length of the normal metal $N_N$. These $0-\pi$-transitions as a function of length are shown in \cref{fig:spin_current_oscillations}, where the transition in strain can also be seen for $N_N=3,6$. Further changes in sign occur at other lengths for higher strain (see Fig. (5) in the main text). 

\section{Comparison of curvatures, strain and SOC-energy}
\setcounter{table}{0} 
We take the strain-induced SOC \cite{salamone2022curvature_sup} 
$\alpha_N = a_N \kappa$ to be of the order $\alpha_N \approx \kappa$. For specificity, we consider a BCC lattice Niobium film with $a=1.65$ Å\cite{gonzalez2021cooper},  $m^* = 2.14 m_e$,  $|k_F| = 1.18$ Å$^{-1}$,\cite{prozorov2022niobium} and the gap $|\Delta| = 2.32$ meV \cite{bonnet1967new}. In \cref{Tab:strain values}, we then compare the SOC energy $\Delta_{\text{soc}} = \hbar^2 \kappa |k_F|/m^*$ with the gap for different values of strain, and show how this relates to the geometric curvature $\kappa=1/R$, where $R$ is the radius of curvature, for three different film thicknesses, $N_n=(3,6,9)$. Note that since the SOC strength scales with the curvature, and the strain scales with the curvature and thickness, the effective SOC strength scales with strain and the inverse of the thickness. This is apparent in Fig. (3) in the main text, where we see a shift along strain-axis for increasing $N_n$.
\vspace{0.5mm}

\begin{table}[H] 
\centering
\begin{tabular}{cccc} 
\toprule
${\ }$ $N_n$ ${\ }$ & ${\ }$ $\kappa$ (nm$^{-1}$) ${\ }$ & ${\ }$ $\epsilon_{ss}$ (\%) ${\ }$ & ${\ }$ $\Delta_{\text{soc}}/\Delta$ ${\ }$\\
\midrule
     & 0.061 & 1 & 10.70  \\ 
3    & 0.121 & 2 & 21.40 \\ 
     & 0.182 & 3 & 32.10  \\ 
\midrule
     & 0.024 & 1 & 4.28  \\ 
6    & 0.048 & 2 & 8.56 \\ 
     & 0.073 & 3 & 12.84 \\ 
\midrule
     & 0.015 & 1 & 2.67 \\ 
9    & 0.030 & 2 & 5.35   \\ 
     & 0.045 & 3 & 8.02\\ 
\bottomrule
\end{tabular}
\caption{In the table above are different thicknesses of Niobium ($N_n$ layers). We give the curvature $\kappa$ and the resulting strain $\epsilon_{ss}$ (\%). In the last column we compare the SOC energy $\Delta_{\text{soc}}$ with the superconducting gap $\Delta$.} \label{Tab:strain values}
\end{table}

\end{document}